\documentclass[conference]{IEEEtran}

\makeatletter

\def\ps@IEEEtitlepagestyle{%
  \def\@oddfoot{\mycopyrightnotice}%
  \def\@evenfoot{}%
}
\def\mycopyrightnotice{%
  \gdef\mycopyrightnotice{}
}

\usepackage{blindtext}
\usepackage{eso-pic}
\IEEEoverridecommandlockouts
\usepackage{cite}
\usepackage{amsmath,amssymb,amsfonts}
\usepackage{algorithmic}
\usepackage{graphicx}
\usepackage{textcomp}
\usepackage{xcolor}

\usepackage{multirow}%
\usepackage{amsmath,amssymb,amsfonts}%
\usepackage{mathrsfs}%
\usepackage{manyfoot}%
\usepackage{booktabs}%
\usepackage{listings}%
\usepackage{caption}
\usepackage{subcaption}
\usepackage{float} 
\usepackage{wrapfig}
\usepackage{array}
\usepackage{textcomp}
\usepackage{tabularray}
\usepackage{url}
\usepackage{longtable}
\usepackage{subcaption}
\usepackage{flushend}

\def\BibTeX{{\rm B\kern-.05em{\sc i\kern-.025em b}\kern-.08em
    T\kern-.1667em\lower.7ex\hbox{E}\kern-.125emX}}
    
\usepackage{eso-pic}

\begin{document}
\title{\vspace*{1cm}
Can You Keep Calm?: \\Adaptive Gameplay using Heart Rate as a Controller\\
}

\author{\IEEEauthorblockN{1\textsuperscript{st} Md Mosharaf Hossan}
\IEEEauthorblockA{\textit{Department of Computer Science} \\
\textit{Idaho State University}\\
Pocatello, United States \\
mdmosharafhossan@isu.edu}
\and
\IEEEauthorblockN{2\textsuperscript{nd}  Rifat Ara Tasnim}
\IEEEauthorblockA{\textit{Department of Computer Science} \\
\textit{Idaho State University}\\
Pocatello, United States \\
rifataratasnim@isu.edu}
\and
\IEEEauthorblockN{3\textsuperscript{rd} Farjana Z Eishita}
\IEEEauthorblockA{\textit{Department of Computer Science} \\
\textit{Idaho State University}\\
Pocatello, United States \\
farjanaeishita@isu.edu}
}

% \IEEEaftertitletext{\vspace{-1.5\baselineskip}}

\maketitle
% \conf{\textit{  V. International Conference on Electrical, Computer and Energy Technologies (ICECET 2025) \\ 
% 3-6 July 2025, Paris-France}}

\begin{abstract}
Serious games for health are designed with specific health objectives and are increasingly being used in mental health interventions. Leveraging sensor equipped handheld devices such as smartphones and smartwatches, these games can provide accessible and engaging therapeutic environments. This study introduces a heart rate (HR) controlled game to aid players manage stress by adjusting gameplay according to their biometric feedback. We aimed to determine how HR-based controls influence their experience and if it can be used to reduce stress. Findings from a controlled experiment revealed that HR controlled gameplay reduced negative and increased positive emotions. Also, players exhibited relatively less cardiac reactivity in HR adaptive target based gameplay. This highlights the promise of biometric feedback based gamified digital environments in supporting accessible mental health support.
\end{abstract}

%\copyrightnotice{XXX-X-XXXX-XXXX-X/XX/\$XX.00 ©20XX IEEE}

% target based gameplay when game elements were controlled based on HR.


\section{Introduction}
Video games have demonstrated their impact beyond `entertainment only' in numerous applications in various fields, including health. With technological advancement, handheld devices equipped with various sensors have enhanced opportunities in serious game research for health. The term `serious game' is used to describe digital games where the goals go beyond the playfulness\cite{charsky2010edutainment}. Researchers have found that playing video games can positively affect cognitive, motivational, emotional, and social domains\cite {granic2014benefits}. Considering this aspect, scientists have applied the concept of serious gaming in various sectors of health, such as detecting and assessing psychological disability, improving the efficacy of existing techniques, and educating and training people to help individuals suffering from mental disorders.

Mandryk et al. \cite{mandryk2017games} mentioned recent research showing how games facilitate stress recovery by rewarding players and down-regulating negative affect. This gamified approach has been used to treat several mental disorders, such as phobias, stress-related disorders, depression, eating disorders, and chronic pain. Through this technology, the therapist has total control over the virtual situations and elements in the computer program. Also, patients feel more secure during therapy because outcomes they fear will happen in the real world cannot happen in the game world\cite{ventura2018virtual}. 

% Video games have a strong capability in mental health applications, both by allowing easier access to treatment and by providing better client engagement\cite{doherty2010design}\cite{wrzesien2011mixing}.

 The integration of biometric feedback into digital games has opened new possibilities to enhance the experience of gameplay and support mental health interventions. Biofeedback techniques integrated into games have been effective in reducing stress and promoting emotional self-regulation \cite{russoniello2009effectiveness} \cite{jafri2021calma}. Among various physiological signals, heart rate (HR) has gained attention due to its direct association with stress, arousal, and emotional states. By dynamically adapting game difficulty based on real-time HR, games can provide personalized experiences that respond to players' physical and emotional conditions. Espinoza et. al. \cite{espinoza2023biocontrolledgame} showed that HR variation can act as an effective biofeedback signal to regulate stress and enhance engagement in adaptive games.

% Adaptive gameplay mechanisms, driven by HR, offer a unique opportunity to investigate how players interact with games under changing difficulty levels. Unlike traditional static games, HR-based adaptive games can adjust their challenges to match players' stress levels to create a more balanced and engaging experience. For example, games that reduce difficulty when elevated HR levels are detected have been shown to promote calmness and improve user engagement \cite{moschovitis2023horrorgamedda}. This study explores whether such adaptations can improve gameplay experiences and influence players' emotional regulation.

In addition, introducing specific targets within adaptive game environments can further shape player behavior and engagement. Achieving a goal or milestone can influence players' sense of accomplishment and emotional response, particularly in games that dynamically adjust difficulty. Patzer et. al. \cite{patzer2020developing} demonstrated that target-driven gameplay can improve motivation and overall satisfaction by providing players with clear objectives and structured challenges. Understanding the role of target-driven gameplay in HR adaptive games is crucial for designing games that are both enjoyable and engaging.

% Also, HR adaptive games have the potential to serve as tools for stress reduction. By monitoring and responding to players' physiological states, these games can provide real-time interventions that help players maintain calmness during challenging scenarios. This creates opportunities for practical applications in stress management and mental health support.

Arcade games are often designed to accommodate infinite gameplay with rapidly increasing difficulty. These games are generally highly addictive in nature \cite{gao2022nature}. In arcade style game design, the focus is often on the reflexes of the players rather than solving puzzles or complex thinking. Carrie J. Cai \cite{cai2013adapting} adapted arcade game Tetris in the context of serious game. Tijs et al. \cite{tijs2009creating} used an adapted version of the arcade game Pacman with variable speed mechanics to identify correlation between game difficulty, players' emotional states, and HR. These characteristics of arcade games provide an opportunity to study how players' experience change if they are adapted based on HR. 

% In addition, this creates opportunities for practical applications in stress management and mental health support.

% Also, if a casual arcade game can be used in stress management.

% They presented a web-based vocabulary drill game augmented with speech recognition to help adult learners acquire new vocabulary. 
% Psychological resilience (PR) \cite{psychological_resilience} refers to the ability to cope with a crisis and to quickly return to precrisis status. Gamified approach to build PR have been studied in prior studies \cite{habibi2023empathetic} \cite{harmon2021magic}. This also provides opportunities to adapt the game in such way that helps players return to calmness after a crisis inside the game and build PR.

In this study, we provide a gamified digital platform inspired by the famous arcade game Flappy Bird~\cite{flappybirdwiki} that adapts it's difficulty based on players' HR change. The research questions include:

% This study aims to answer the following research questions:

\begin{enumerate}
    
    \item How HR-based difficulty adaptation changes Player Experience (PX)?

    \item Is there a difference in PX when there is a target to achieve in an adaptive game environment?

    \item Can HR adaptive games aid in reducing stress?

\end{enumerate}

Through a controlled experiment, we ascertained that HR adaptive game levels 
enhanced PX. Also, players showed less cardiac reactivity (CR) and felt more positive emotions with target-driven gameplay. The findings aim to provide new insights into the design of adaptive games that not only entertain but also support emotional well-being.

% helped players stay calmer while enhancing their gameplay experience.
% heart rate modulated levels enhanced gameplay experience and . Also, . 

\section{Related Work}
% \cite{espinoza2023biocontrolledgame} investigated biofeedback controlled games that utilize HR to assist players in regulating their emotional states. Their study highlighted the role of HR variation in creating adaptive experiences that promote calmness and emotional balance.
Heart health as a tool for adaptive game design has been explored extensively in prior research. Espinoza et al. \cite{espinoza2023biocontrolledgame} explored how to introduce biofeedback in short casual video games as a way to assist players in regulating their emotional states. The game changes its difficulty based on the player's heart rate variability (HRV) and allows players to navigate through different emotional states. Their study highlighted the role of HRV in creating adaptive experiences that promote calmness and emotional balance. Similarly, Parnandi and Gutierrez-Osuna \cite{parnandi2017visual} demonstrated that HRV-based adaptations in games can effectively transfer relaxation skills to real-world scenarios. Jafri et al. \cite{jafri2021calma} developed an HRV-based adaptive mobile and smartwatch game that helps players self-regulate anxiety through breathing exercises. However, these studies use HRV to measure players' emotional states, and accurate HRV measurement requires specialized equipment that is often costly and inaccessible to the general population. To make our game widely accessible, we looked at the availability of devices. Portable devices such as smartwatches are widely available. However, smartwatches lack the precision required for reliable HRV measurement \cite{dobbs2019accuracy}. HR, on the other hand, can be measured with considerable accuracy using smartwatches \cite{spinsante2019accuracy} \cite{phan2015smartwatch} \cite{nissen2022heart}. Additionally, HR changes serve as a robust indicator of stress \cite{harmon2021magic} \cite{taelman2009influence} \cite{trotman2019associations}. For this reason, we selected HR as a practical and effective biometric indicator for our adaptive design.

% Many studies utilize Heart Rate Variability (HRV) to assess players' emotional states. However, 

% However, these studies use HRV to measure players' emotional states. Gadgets for accurate HRV measurement is expensive and not available to the mass population. Portable devices such as smartwaches are available to the general public. However, HRV cannot be measure accuratly with those devices \cite{dobbs2019accuracy}. HR, on the other hand, can be measured quite accurately through smartwatches \cite{spinsante2019accuracy} \cite{phan2015smartwatch} \cite{nissen2022heart}. Along with HRV, HR change is also one of the indicators of stress \cite{harmon2021magic} \cite{taelman2009influence} \cite{trotman2019associations}. For this reason, HR was chosen as an effective bio indicator in our adaptive design. 

Larkin et al. \cite{larkin1992effects} integrated HR biofeedback with score contingency (SC) in a game designed to decrease CR. CR is defined as the average HR increase in response to stressors. The score of the game was shown on a display. The study aimed to assess skill transfer to novel tasks (e.g., mental arithmetic). In a 2×2 design, SC significantly reduced HR reactivity during gameplay and mental arithmetic. A similar route was taken by Moschovitis et al. \cite{moschovitis2023horrorgamedda}, who introduced `Caroline', a horror game that adjusts its difficulty based on players' HR. It encourages calmness by increasing the difficulty of the game when players are stressed and decreasing when players are calm. This approach increased player motivation compared to a non-adaptive version of the game. Bevilacqua et al. \cite{bevilacqua2018changes}  explored the connection between physiological (HR) and behavioral (facial actions) indicators and emotional states in gaming sessions designed to provoke boredom and stress. They used a custom game with linearly increasing difficulty. The study demonstrated significant HR and FA variations between these emotional states. Tijs et al. \cite{tijs2009creating} investigated the role of physiological and emotional feedback in creating emotionally adaptive games. Using an adapted version of the arcade game Pacman with variable speed mechanics, they identified correlations between game difficulty, self-reported emotional states, and physiological indicators such as HR and skin conductance. Their study demonstrates the potential for real-time game adaptation based on physiological data. Inspired by these studies, we looked at HR through the CR of players during different levels of our game to understand if HR adaptive levels can help players reduce stress.

Some prior studies have also investigated using biofeedback to adapt game elements accordingly to improve the overall play experience. Mirza-babaei et al. \cite{mirza2011understanding} demonstrated the importance of biofeedback to improve the PX. They changed horror elements in Half-Life 2 based on player biometric data and found that it improved the PX. Hossan et al. \cite{hossan2024adaptive} provided a list of studies that used various physiological signals to adapt game elements. Signals such as electrodermal activity, HR, and facial expressions were discussed. In their more recent study, Hossan et al. \cite{hossan2025gamified} discuss the possibility of integrating biofeedback for real-time emotion regulation in gamified exposure therapy sessions. Nenonen et al. \cite{nenonen2007hrinteractivegame} presented a novel way of using real-time HR to control a physically interactive biathlon (skiing and shooting) computer game. They found that this approach promoted exercise through gameplay. Another study that focused on the increase in physical activity of players through enhanced gaming experience was done by Magielse et al. \cite{magielse2009heartbeat}. They presented `HeartBeat', a pervasive game designed for children that uses their HR to promote physical activity and social interaction. Frachi et al. \cite{frachi2022design} designed a two-dimensional (2D) storytelling game where the game mechanics and interactions are based on players' emotions, and they reported positive reception of the approach. In their more recent study, Frachi et al. \cite{frachi2023adaptivespeed} investigated how controlling some of the game mechanics using biofeedback affects physiological reactions, performance, and the experience of the player. More specifically, they assess how different game speeds affect players' physiological responses and game performance. However, these studies focused solely on improving PX. Our study aims to enhance the gameplay experience while helping them reduce stress during gameplay.

These studies collectively establish the foundation for exploring HR-driven adaptive games. Building on this body of work, our study aims to investigate how HR-based difficulty adaptation and target-driven gameplay can enhance engagement and support stress reduction.

\section{Methodology}

\subsection{Our Approach}

The purpose of this study was to investigate players' perceived gameplay experience when an arcade style casual game is played with and without HR-based adaptation. A handheld game imitating the renowned Flappy Bird \cite{flappybirdwiki}  was developed from scratch. The game had three different levels of difficulties (details in section~\ref{sec:game}). A smartwatch (Google Pixel watch) was used to read and transmit real-time HR data to a server. An AWS ec2 instance was used as a server to facilitate the near real-time HR transfer between a smartwatch and an Android phone.

\subsection{The Game} \label{sec:game}

Inspired by the Flappy Bird~\cite{flappybirdwiki}, our game features a bird that the player navigates through gaps of right-to-left moving inverted pillars by tapping on the screen. Each tap makes the bird fly vertically; otherwise, it falls due to gravity. The goal is to time the taps meticulously so that the bird stays inside the opening when a set of inverted pillars arrives. If a bird touches a pillar, it is pushed to the left. Repeated touch can push it off-screen, resulting in `game over'. Similarly, `game over' can occur if the bird flies or drops out of the screen due to frequent touch or no touch at all. 

%Ma'am, should this be added here
After a player makes a mistake by hitting a pillar, the pillar drags the bird to the left instead of ending the game like the original Flappy Bird~\cite{flappybirdwiki}. The player might feel stressed as only a few mistakes are left before being dragged out of the screen. This stress response is expected to increase Player HR. We reduce the difficulty by decreasing pillar speed if the player's HR goes too high. We provide the chance to continue playing the game as part of our effort to see if reduced difficulty inside the game based on stress response helps players enjoy the game even more and motivates them to play longer. The game features three levels with unique constraints. 

% Also, if this strategy can help players lower subsequent stress response. 

\subsubsection{Level 1}

\begin{figure}[H] % Use H to force the figure to appear exactly here
    \centering
    % \vspace{-5pt} % Adjust as needed to avoid overlap with preceding text
    \includegraphics[width=0.43\textwidth]{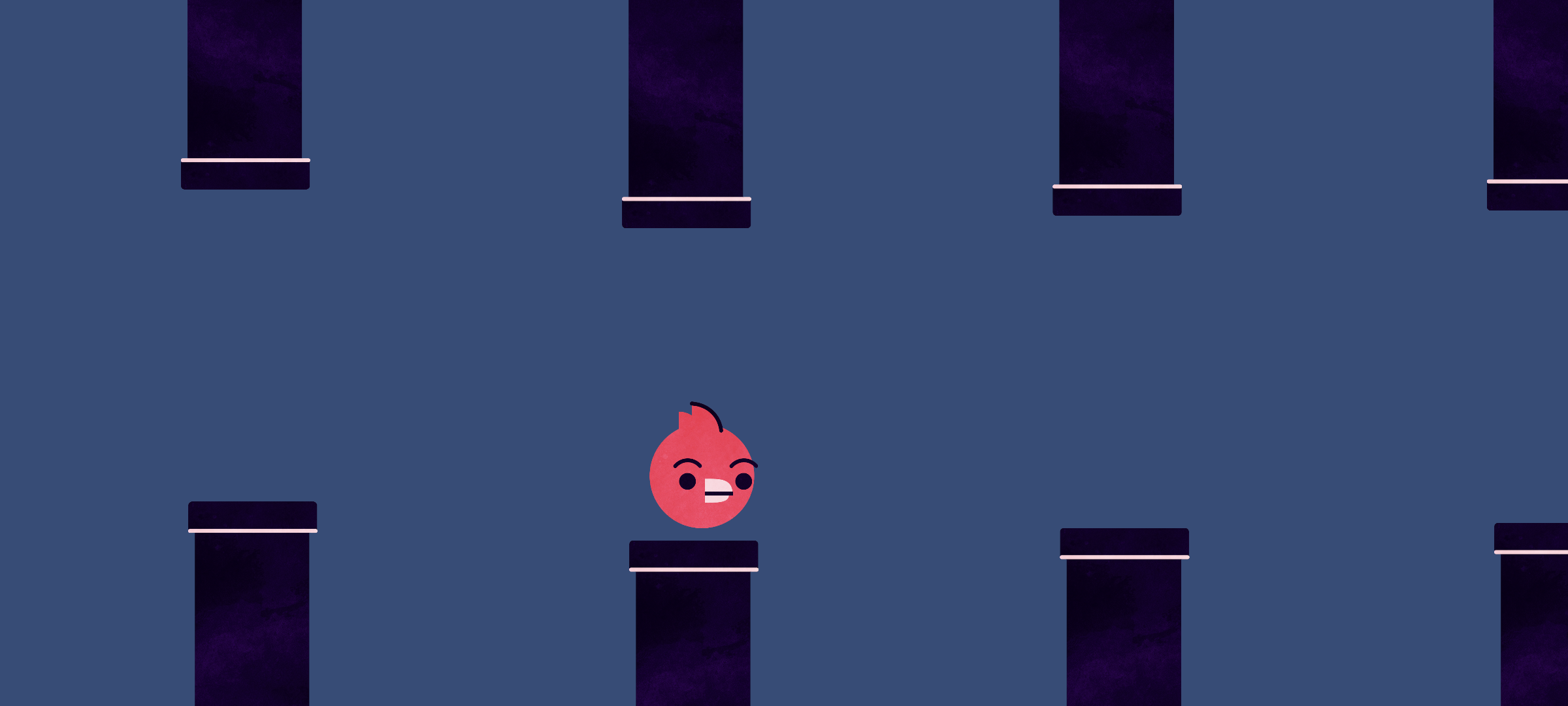} % Set the width as needed
    \caption{Level 1 gameplay}
    \label{fig:Level1Gameplay} % Ensure the label has no spaces or special characters
    % \vspace{-20pt} % Adjust as needed to control spacing after the figure
\end{figure}

As shown in Figure \ref{fig:Level1Gameplay}, this level mimicked simple gameplay with no scoring feedback or HR-based control. The first pillar starts moving at an initial speed. Subsequent pillars increase speed by 10\%, and the gap between them increases by 20\% of the initial gap. This increase in constraints keeps happening until a limit is reached to keep the game playable.

% \begin{wrapfigure}{r}{0.5\textwidth} % 'r' for right, '0.4\textwidth' for width of the wrap
%     \centering
%     \vspace{-5pt} % Adjust as needed to avoid overlap with preceding text
%     \includegraphics[width=\linewidth]{fb_without_score.png}
%     \caption{Level 1 gameplay}
%     \label{fig:Level1Gameplay} % Ensure the label has no spaces or special characters
%     \vspace{-20pt} % Adjust as needed to control spacing after the figure
% \end{wrapfigure}

%Mosharaf - The last sentence lacks clarity. Please re-write
% Addressed
\subsubsection{Level 2}

The pillars start moving similarly to level 1 in this level. This level shows the number of pillars crossed as a score and an HR threshold. Base HR is measured with the average of the first 5 HR logged before the start of gameplay. Then, a pivot value of +5 is added to the base HR to act as a threshold. During gameplay, if the player's HR goes over the threshold, the speed of moving pillars is reduced by 30\%. The goal was to select the reduction rate in a level so that the players could comprehend the difference without hindering their overall flow. A nuanced change or an abrupt reduction might fail to help players calm down. The pivot value of +5 was chosen as an experimental value. If a pivot value is too small, it might not reflect cognitive stress. On the other hand, a pivot that is too big might not provide a speed reduction opportunity before the game ends.

\subsubsection{Level 3}

This level adds a target score of 30 as a constraint on top of the previous features of Level 1 and Level 2. A score of 30 was chosen to develop the game as a short casual arcade game. It takes $\approx$ 1 minute to reach the score 30. The time can vary based on how many times the threshold was breached and the speed of the pillars decreased.

% \subsubsection{Threshold Calculation}
% Before the start of the gameplay, the player's base HR is measured with an average of the first 5 HR logged. Then, a pivot value of +5 is added to that to act as threshold.

\subsection{Data Collection Method}

The Positive and Negative Affect Schedule (PANAS) \cite{watson1988development} and Player Experience Inventory (PXI) \cite{pxiquestionnaire} were used to record the experience of players after each session of play. PANAS was used to log players' positive and negative experiences of the gameplay. It is a self-report questionnaire consisting of two 10-item scales to capture players' positive and negative emotional states. Each item is rated on a 5-point scale of 1 (not at all) to 5 (very much).

The player experience inventory (PXI) \cite{pxiquestionnaire} was used to log how players experienced each level of the game functionally and emotionally. Items across both surveys were measured on a 7-point Likert scale, ranging from -3-Strongly disagree to 3-Strongly agree. The questionnaire consists of 33 questions relating to different constructs:

% % \vspace{1pt}
% \begin{itemize}
%   \item Functional Consequences: Ease of Control (3 items), Challenge \break (3 items), Progress Feedback (3 items), Goals and Rules(3 items), Audio-visual Appeal (3 items)
  
%   \item Psychosocial Consequences: Meaning (3 items), Immersion (3 items), Mastery (3 items), Curiosity (3 items), Autonomy (3 items)
% \end{itemize}

\begin{itemize}
    \item \textbf{Functional Consequences:} Ease of Control (3 items), Challenge (3 items), Progress Feedback (3 items), Goals and Rules (3 items), Audio-visual Appeal (3 items)
    \item \textbf{Psychosocial Consequences:} Meaning (3 items), Immersion (3 items), Mastery (3 items), Curiosity (3 items), Autonomy (3 items)
\end{itemize}

% \begin{itemize}
%     \item \textbf{Functional Consequences:}
%     \begin{itemize}
%         \item Ease of Control (3 items)
%         \item Challenge (3 items)
%         \item Progress Feedback (3 items)
%         \item Goals and Rules (3 items)
%         \item Audio-visual Appeal (3 items)
%     \end{itemize}
%     \item \textbf{Psychosocial Consequences:}
%     \begin{itemize}
%         \item Meaning (3 items)
%         \item Immersion (3 items)
%         \item Mastery (3 items)
%         \item Curiosity (3 items)
%         \item Autonomy (3 items)
%     \end{itemize}
% \end{itemize}

\section{Experiment}

A controlled experiment was conducted on 25 participants (18 male and 7 female). Participants were recruited through email invitations. Their age range along with the respective number of individuals: 15-20 years (1 player), 21-25 years (13 players), 26-30 years (4 players), 31-35 years (2 players), and 35-40 years (5 players). The experiment was conducted under the oversight of an institutional review board approval named `Gamified Digital Intervention to Enhance the Efficacy of Exposure Therapy for OCD'. Each participant signed a consent letter, and a small sum of honorarium was given after gameplay. The game was played on a Google Pixel 7a with a full-screen 6.1-inch display in a dedicated lab. They were prohibited from engaging in distractions during gameplay. Participants played 3 sessions of each level and completed questionnaires afterward. The independent variable is the game level, with three conditions: Level 1, Level 2, and Level 3. The dependent variables include players' HR, PANAS, and PXI. The experiment was designed in a within-subject fashion. The sessions were ordered using the Latin Square Method~\cite{keedwell2015latin} to avoid learning effect bias.

% This experiment served as a proof to understand how players perceive HR for emotion based adaption. We want to use this knowledge in our later experiment where we try to perform digital exposure therapy and use HR as a stress indicator.
% and one level 1 game session was used for game environment familiarization training.
\section{Result} 

% This section includes PANAS and PXI questionnaire responses, summaries of players' average HR during gameplay, and CR of players. To mitigate learner bias, only HR data from the second and third session was used in the analysis. Questionnaire responses from the three levels were analyzed using repeated measures ANOVA for PANAS positive/negative affect, PXI constructs, CR. Because we were comparing more than two repeated conditions (levels) within the same participants, a repeated-measures ANOVA was the most appropriate statistical approach to test for overall differences and control for within-subject variability.

This section includes PANAS and PXI questionnaire responses, summaries of players' average HR during gameplay, and CR of players. HR data from the first session was discarded to reduce the influence of learner bias. Since players played the sessions without breaks, HR data from the second and third sessions were considered a single stressor for calculating CR. This allowed more profound insights into how players' HR reacted over two gameplay sessions. Questionnaire responses from the three levels were analyzed using repeated measures ANOVA for PANAS positive/negative affect, PXI constructs, and CR. Because we were comparing more than two repeated conditions (levels) within the same participants, a repeated-measures ANOVA was the most appropriate statistical approach to test for overall differences and control for within-subject variability.

\subsection{Positive and Negative Affect Schedule (PANAS)}

% A single-factor repeated measures ANOVA on positive affect (PA) yielded 
% \(\textit{F = 4.49, p = 0.016}\). This suggests players experienced significantly different levels of positive and negative emotions when playing these levels. 

\begin{figure}[H] % Use H to place the figure exactly here
  \centering
  \begin{subfigure}[b]{0.4\textwidth} % Adjust width as needed
    \centering
    \includegraphics[width=\textwidth]{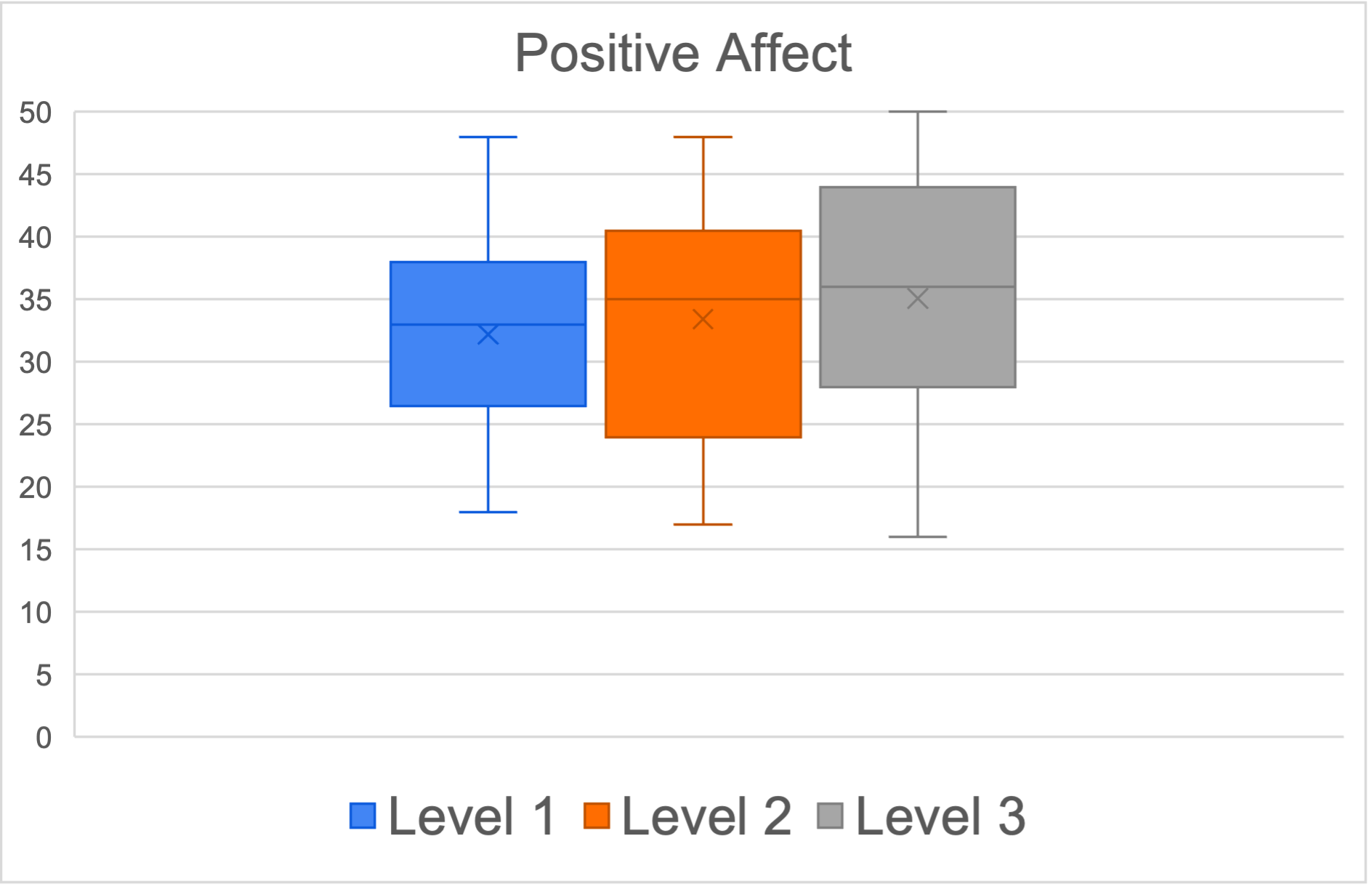}
    \caption{Positive Affect}
    \label{fig:PANAS_positive_affect}
  \end{subfigure}
  
  \vspace{5pt} % Adjust the vertical gap as needed
  
  \begin{subfigure}[b]{0.4\textwidth} % Adjust width as needed
    \centering
    \includegraphics[width=\textwidth]{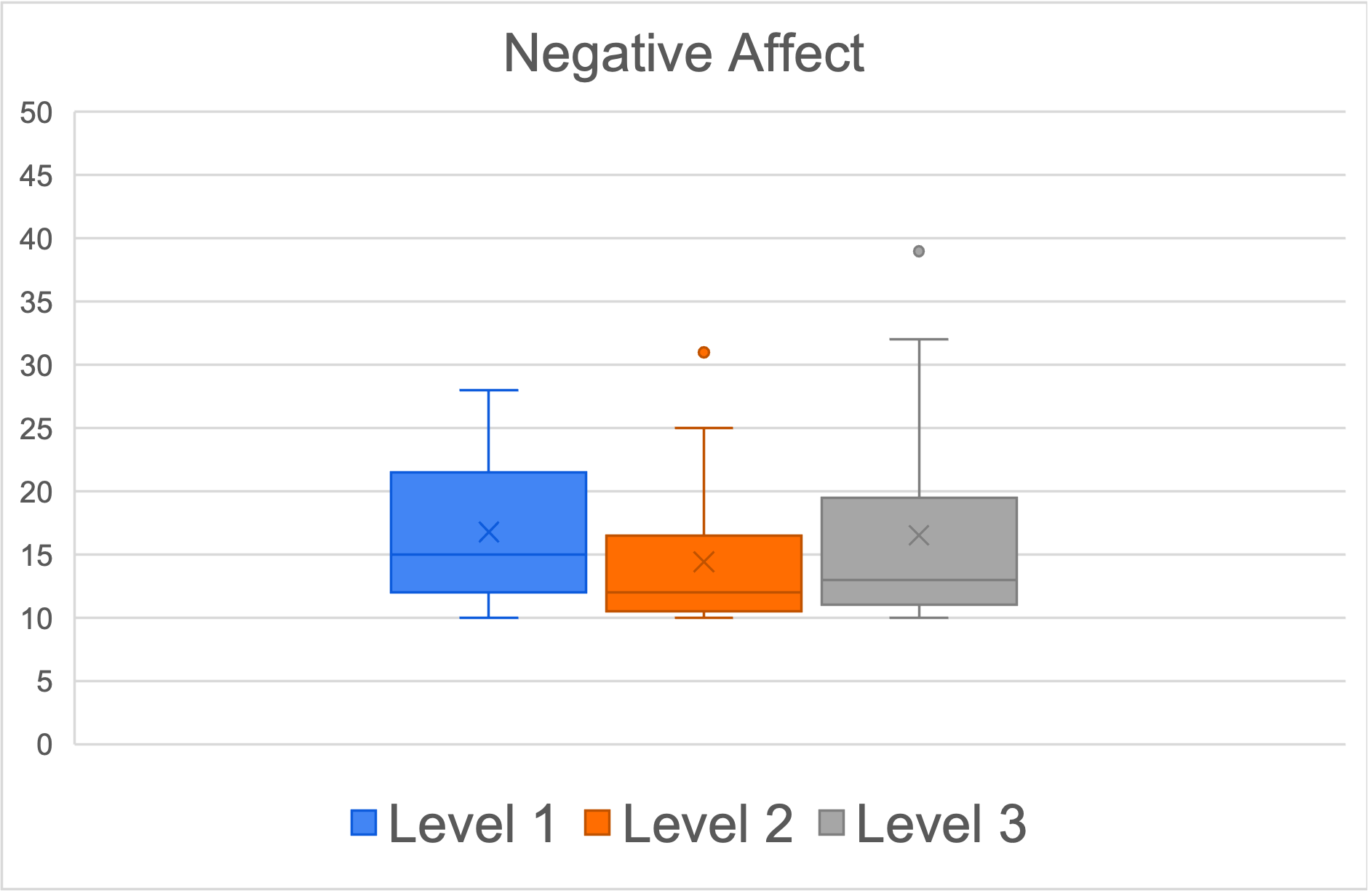}
    \caption{Negative Affect}
    \label{fig:PANAS_negative_affect}
  \end{subfigure}
  
  \caption{PANAS Positive and Negative Affect}
  \label{fig:PANAS_affect}
\end{figure}

Figure \ref{fig:PANAS_affect} shows boxplots of players' PANAS responses. A repeated measures ANOVA ($\alpha$ = 0.05) showed a significant effect of Positive Affect on the measure, (\(F = 4.491, p = .016\)).

% Bonferroni-corrected pairwise comparisons indicated a significant difference between Level 1 and Level 3 (\(p = .021\)), while Level 1 vs Level 2 and Level 2 vs Level 3 were not significant.

% \textbf{A post hoc analysis with pairwise comparison with bonferroni correction showed Level 3 differs significantly from Level 1 with PA = 0.02 and NA = }

% \begin{figure}[H] % Use H to place the figure exactly here
%   \centering
%   \includegraphics[width=0.37\textwidth]{positive_affect.png}
%   \caption{Positive Affect}
%   \label{fig:PANAS_positive_affect}
% \end{figure}
% \vspace{-.5cm}
% \begin{figure}[H] % Use H to place the figure exactly here
%   \centering
%   \includegraphics[width=0.37\textwidth]{negative_affect.png}
%   \caption{Negative Affect}
%   \label{fig:PANAS_negative_affect}
% \end{figure}

% \begin{figure*}[t]
%   \centering
%   \begin{subfigure}[b]{0.45\textwidth}
%     \centering
%     \includegraphics[width=\textwidth]{positive_affect.png}
%     \caption{Positive Affect}
%   \end{subfigure}
%   \hfill
%   \begin{subfigure}[b]{0.45\textwidth}
%     \centering
%     \includegraphics[width=\textwidth]{negative_affect.png}
%     \caption{Negative Affect}
%   \end{subfigure}
%   \caption{PANAS Positive and Negative Affect}
% \end{figure*}

A repeated measures ANOVA ($\alpha$ = 0.05) on Negative Affect revealed a significant effect of Level on the measure, (\(F = 3.87, p = .041\)).

% Bonferroni-corrected pairwise comparisons indicated that Level 2 differed significantly from Level 1 (\(p = .044\)) and from Level 3 (\(p = .017\)), whereas Level 1 and Level 3 did not significantly differ (\(p = 1.00\)).

\subsection{Player Experience Inventory (PXI)}

\begin{figure}[H] % Use H to place the figure exactly here
  \centering
  \begin{subfigure}[b]{0.37\textwidth} % Adjust width as needed
    \centering
    \includegraphics[width=\textwidth]{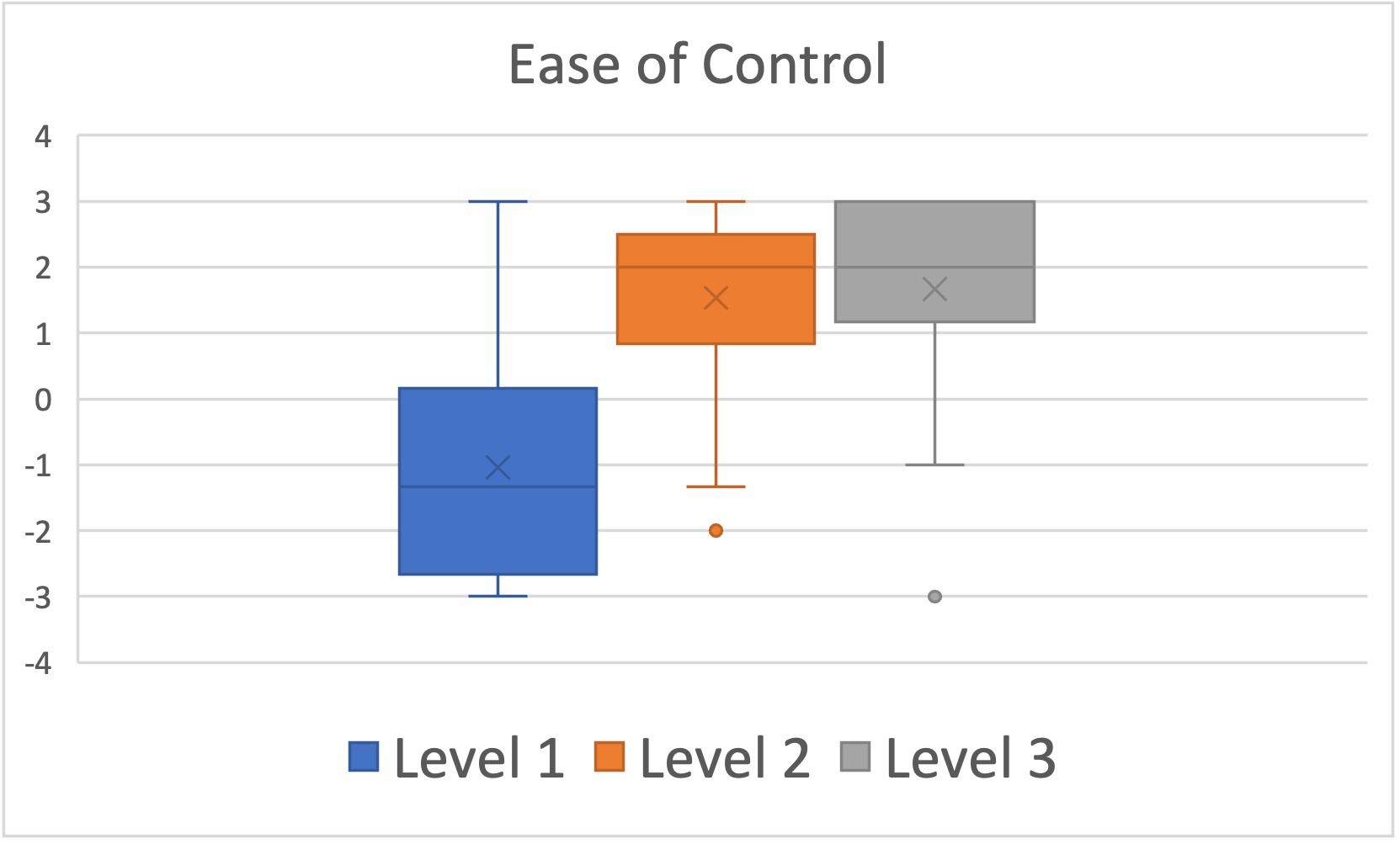}
    \caption{Ease of control}
    \label{fig:Ease_of_control}
  \end{subfigure}
  
  \vspace{6pt}
  
  \begin{subfigure}[b]{0.37\textwidth} % Adjust width as needed
    \centering
    \includegraphics[width=\textwidth]{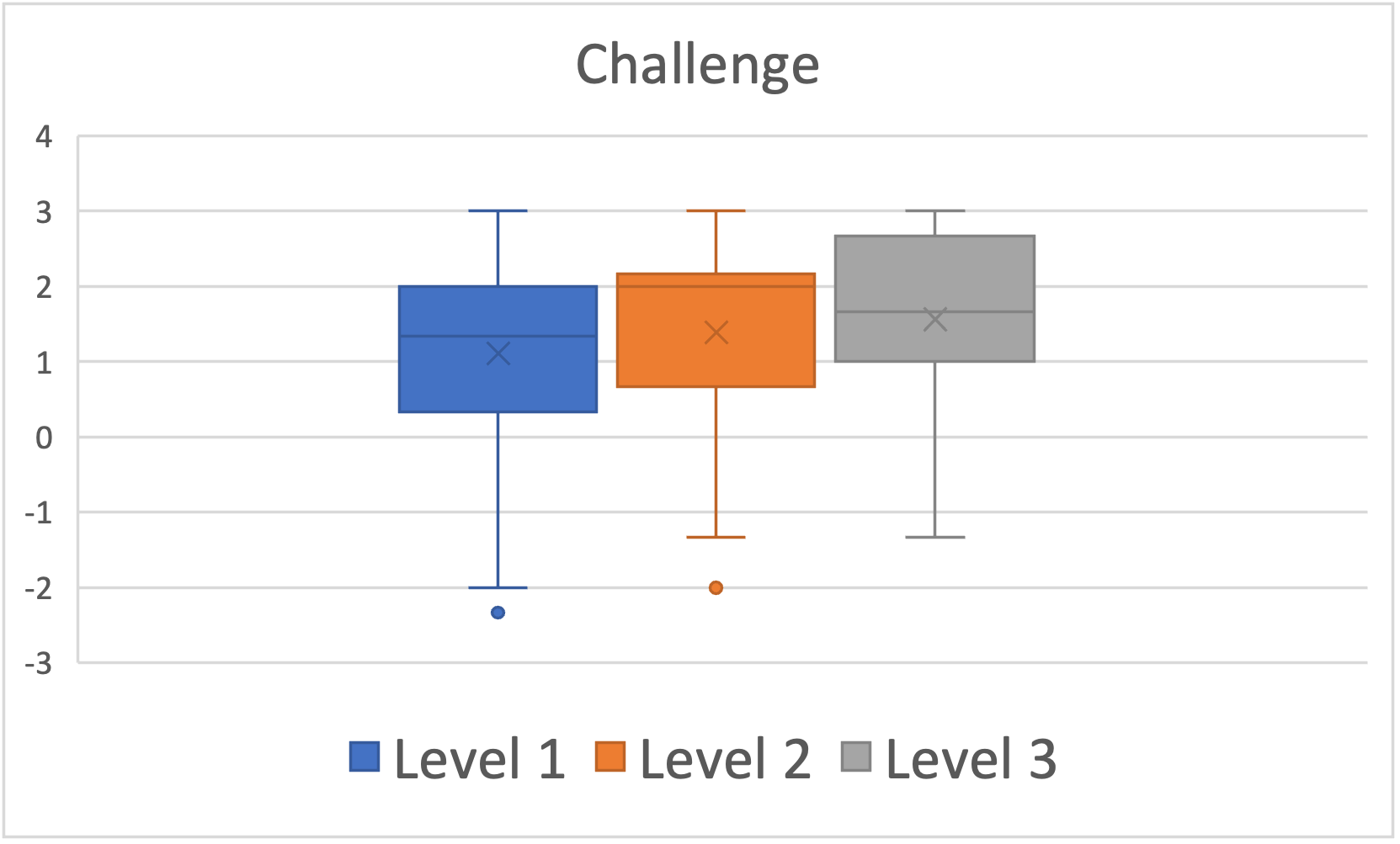}
    \caption{Challenge}
    \label{fig:Challenge}
  \end{subfigure}
  
  \vspace{6pt}
  
  \begin{subfigure}[b]{0.37\textwidth} % Adjust width as needed
    \centering
    \includegraphics[width=\textwidth]{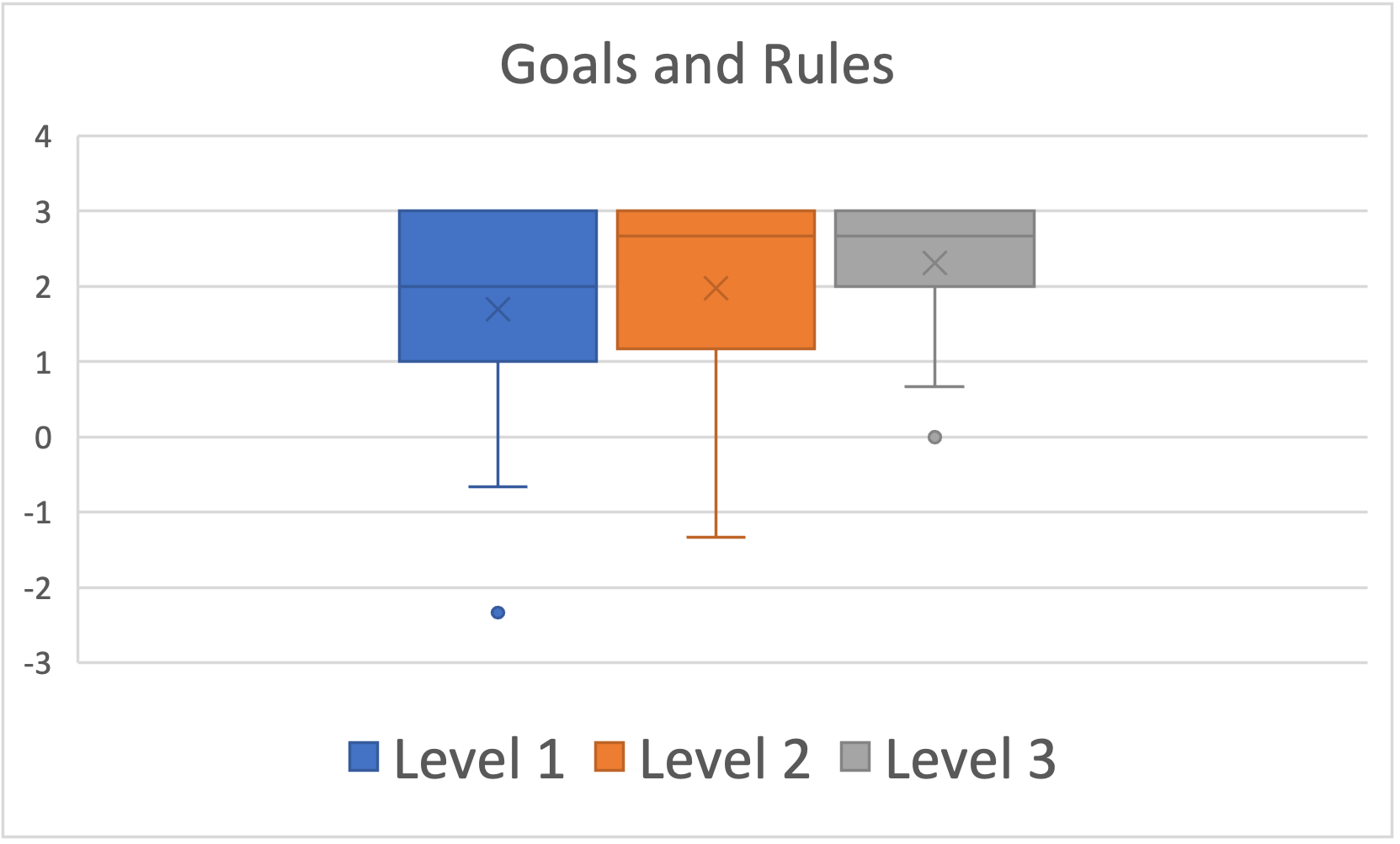}
    \caption{Goals and rules}
    \label{fig:Goals_and_rules}
  \end{subfigure}
  \caption{functional consequences of PXI}
  \label{fig:pxi_functional}
\end{figure}

Table \ref{table:functionalconsequence} and \ref{table:psychosocialconsequence} shows all the \(F\) and \(p\)  values based on PXI constructs.
As shown in Figure \ref{fig:pxi_functional}, the results in functional consequences indicate that players had significantly different experiences regarding ease of control (\(F = 32.765, p \textless 0.001\)), challenge (\(F = 4.937, p = 0.011 \)), and goals and rules (\(F = 3.317, p = 0.044\)). Figure \ref{fig:Ease_of_control} shows players found the latter levels easier to control due to speed reduction, which is consistent with the game's design. As shown in Figure \ref{fig:Challenge}, differences in challenge suggest that HR controlled levels better matched players' skill levels. 

\begin{figure}[H] % Use H to place the figure exactly here
  \centering
  \begin{subfigure}[b]{0.37\textwidth} % Adjust width as needed
    \centering
    \includegraphics[width=\textwidth]{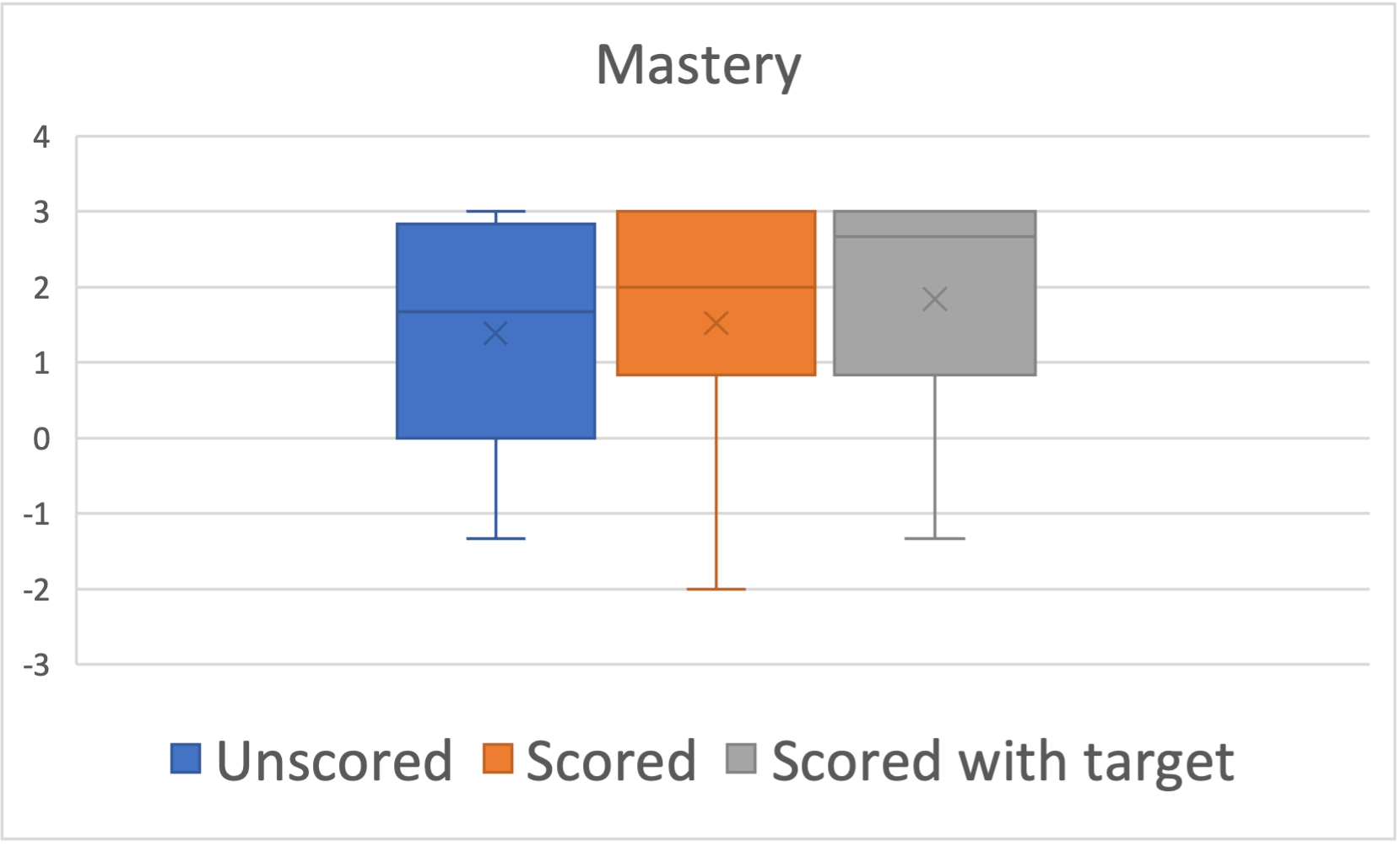}
    \caption{Mastery}
    \label{fig:mastery}
  \end{subfigure}
  \caption{psychosocial consequences of PXI}
  \label{fig:pxi_psychosocial}
\end{figure}

As shown in Figure \ref{fig:pxi_psychosocial}, no significant differences were found in any psychosocial constructs of the PXI questionnaire. However, The analysis of mastery yielded (\(F = 2.64, p = 0.08\)) , suggesting a trend towards significance.

 % \vspace{5pt}

% \begin{figure}[H] % Use H to place the figure exactly here
%   \centering
%   \includegraphics[width=0.4\textwidth]{mastery.png} % Adjust height as needed
%   \caption{Mastery}
%   \label{fig:mastery}
% \end{figure}

% \begin{table*}[htbp]
% \centering
% \caption{PXI Functional Consequences}
% \label{table:functionalconsequence}
% \begin{tabular}{|p{0.8cm}|p{2.2cm}|p{1.65cm}|p{1.65cm}|p{1.65cm}|p{1.65cm}|} % Adjusted column widths for consistency
% \hline
%    & Ease of Control & Challenge & Progress Feedback & Audiovisual Appeal & Clarity of Goals \\
% \hline
%    F  & 32.765 & 4.937 & 0.91 & 2.48 & 3.317 \\
% \hline
%    p & 0.00000000106 & 0.011 & 0.4 & 0.09 & 0.044 \\
% \hline
% \end{tabular}

% \end{table*}

% \begin{table*}[htbp]
% \centering
% \caption{PXI Psychosocial Consequences}
% \label{table:psychosocialconsequence}
% \begin{tabular}{|p{0.8cm}|p{2.2cm}|p{1.65cm}|p{1.65cm}|p{1.65cm}|p{1.65cm}|} % Same column widths as the first table
% \hline
%    & Meaning    & Mastery    & Immersion    & Curiosity    & Autonomy \\
% \hline
%    F & 0.85 & 2.64 & 0.21 & 0.73 & 0.22 \\
% \hline
%    p & 0.43 & 0.08 & 0.8 & 0.49 & 0.79 \\
% \hline
% \end{tabular}

% \end{table*}

\begin{table}[htbp]
\centering
\caption{PXI Functional Consequences}
\label{table:functionalconsequence}
\begin{tabular}{|p{0.5cm}|p{1.1cm}|p{1.1cm}|p{1.1cm}|p{1.2cm}|p{1.1cm}|} % Adjusted column widths for consistency
\hline
   & Ease of Control & Challenge & Progress Feedback & Audiovisual Appeal & Goals and Rules \\
\hline
   F  & 32.765 & 4.937 & 0.91 & 2.48 & 3.317 \\
\hline
   p & $<$ 0.0001 & 0.011 & 0.4 & 0.09 & 0.044 \\
\hline
\end{tabular}
\end{table}
% 0.00000000106

% \vspace{5pt}
% \vspace{-0.2\baselineskip}  % Reduce vertical space here

\begin{table}[htbp]
\centering
\caption{PXI Psychosocial Consequences}
\label{table:psychosocialconsequence}
\begin{tabular}{|p{0.5cm}|p{1.1cm}|p{1.1cm}|p{1.1cm}|p{1.2cm}|p{1.1cm}|} % Same column widths as the first table
\hline
   & Meaning    & Mastery    & Immersion    & Curiosity    & Autonomy \\
\hline
   F & 0.85 & 2.64 & 0.21 & 0.73 & 0.22 \\
\hline
   p & 0.43 & 0.08 & 0.8 & 0.49 & 0.79 \\
\hline
\end{tabular}
\end{table}

% \vspace{-0.3\baselineskip} 
% \subsection{Average HR between levels}
% Average HR were calculated during a gameplay session, and a single factor repeated measure ANOVA was between HR of three game levels. The result demonstrates a borderline significance (\textit{F = 3.161, p=0.0513}).

% \begin{figure}[H] % Use H to place the figure exactly here
%     \centering
%     % \vspace{-10pt} % Adjust as needed to avoid overlap with preceding text
%     \includegraphics[width=0.5\textwidth]{average_heart_rate.png} % Set the width as needed
%     \caption{Average heart rate}
%     \label{fig: Average heart rate} % Ensure the label has no spaces or special characters
% \end{figure}

\subsection{Cardiac Reactivity}
\begin{figure}[H] % Use H to place the figure exactly here
    \centering
    % \vspace{-10pt} % Adjust as needed to avoid overlap with preceding text
    \includegraphics[width=0.43\textwidth]{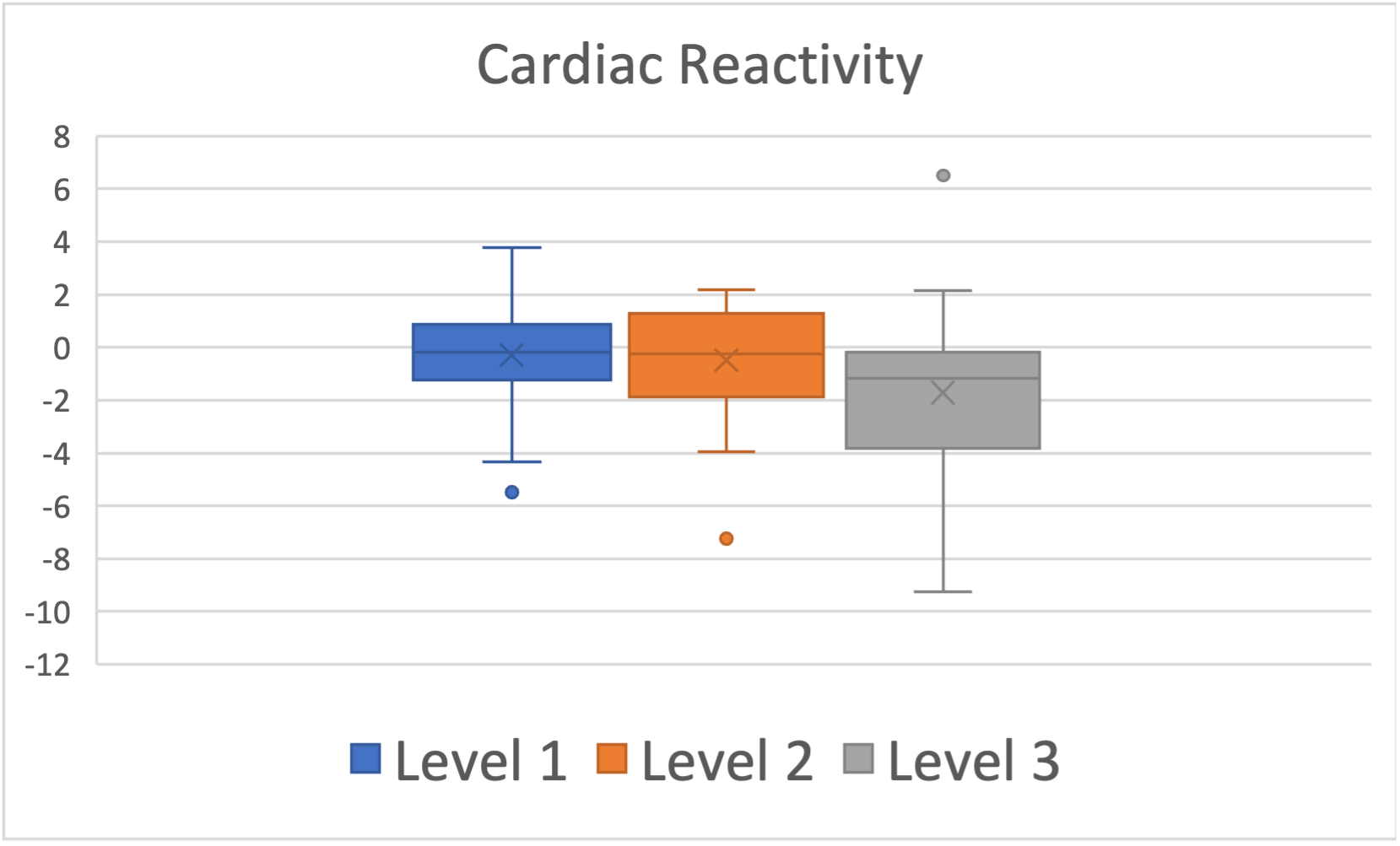} % Set the width as needed
    \caption{Cardiac Reactivity}
    \label{fig: cardiac reactivity} % Ensure the label has no spaces or special characters
\end{figure}

CR was calculated with heart rates documented during second and third gameplay sessions. A single factor repeated measure ANOVA was calculated between CR of three game levels. The result demonstrates a significance of (\(F = 2.85, p=0.067\)). Average gameplay time for each levels were 48, 56, and 51 seconds. Figure \ref{fig: cardiac reactivity} suggests players HR dropped more during gameplay during level 3 compared to other levels despite having similar gameplay times.

% had less average heart rate during gameplay than they started with

% level 3 was able to calm players more within 2 sessions of gameplay.
 
\section{Analysis \& Discussion}

The primary objective of this research is to investigate PX while controlling the in-game action via real-time HR. Three levels of gamplay were developed to analyze the differential impact of players' HR, which eventually controls their in-game performance. The PANAS and PXI data were logged via survey after 3 gameplay sessions of each level.

Both positive and negative affect demonstrates significant differences in experience. Figure~\ref{fig:PANAS_positive_affect} demonstrates that players enjoyed heart rate adaptive level 3 gameplay the most. The implication could be that with this certain genre of game, players like to meet a target. However, the least negative experience was observed in heart rate adaptive level 2 which featured an open gameplay with scoring feedback.  

% There was a significant difference between level 1 and level 3 (\(p = 0.021\)). 

% Level 2 differed significantly from level 1 (\(p = 0.044\)) and level (\(p = 0.017\)).

Player Experience Inventory (PXI) Bench analysis was conducted to analyze players' Functional and Psychosocial Consequences. The Functional Consequences align directly with the players' immediate decisions made based on the game design. Significant difference was found in challenge construct with (\(F = 4.937, p = 0.011 \)). PXI defines challenge as the extent to which the specific challenges in the game match the player's skill level. Players felt appropriately challenged when the game was played with an open-end target with a reduced speed controlled by their real-time HR. This finding is quite aligned with the findings of PANAS negative affect, where people experience the least negative emotion in HR adaptive levels.

% In addition, as a natural implication, players' HR was highest while facing the most challenge in gameplay in level 2 (Figure~\ref{fig: Average heart rate}). 

% This implication is quite aligned with the average gameplay time recorded for level 2 gameplay as well (Figure~\ref{fig:Average gameplay time}).

 Players found level 1 (control group) the most difficult to control. This outcome is consistent with how the game is designed as there is no adaptivity in level 1. 
 
 % One implication of this thought can be that players enjoyed the speed to be controlled by their HR rather than the auto increment of the speed while proceeding forward in the gameplay.

No significant difference was observed in the players' psychosocial consequences indicating that although the different game mechanics left an impact in controlling the game functionally, second-order emotional experience remained similar. However, as medians in Figure \ref{fig:mastery} suggests, players felt a sense of competence and mastery derived from playing the HR controlled levels of the game. While the p-value of 0.08 indicates the absence of strong statistical significance, the observed trend warrants further investigation.

% Player response through questionnaires and CR suggest that difficulty adaptation based on HR has the potential to enhance gameplay experience. However, while target based HR adaptive gameplay has the potential to surge positive emotions, it also has the potential to escalate negative emotions, possibly due to not achieving the target at the very end, as discussed as the peak-end effect \cite{gutwin2016peak}. To answer RQ3, players' average HR reactivity reduced during level 3 gameplay. However, during level 2 the HR reactivity stayed similar to level 1. Further investigation is needed to understand if HR adaptivity itself is able to reduce cardiac reactivity in players.

Player responses through questionnaires and CR suggest that difficulty adaptation based on HR has the potential to enhance the gameplay experience. However, while target based HR adaptive gameplay has the potential to surge positive emotions, it also has the potential to escalate negative emotions, possibly due to not achieving the target at the very end, as discussed as the peak-end effect \cite{gutwin2016peak}. To answer RQ3, players' CR was lower than other levels during level 3 gameplay. However, during level 2, the CR stayed similar to level 1. Further investigation is needed to understand if HR adaptivity itself is able to reduce CR in players.

\paragraph{Threats to Validity} The experiment was conducted with a small sample size that focused only on university students with a particular age range. Hence, a possibility of bias on the outcome remains. Also, an arbitrary threshold value for heart rate increase during cognitive stress introduces bias. However, the value was chosen arbitrarily to act as a reference for future studies. Furthermore, a larger target score in level 3 can provide deeper insights in target based gameplay experience. 

% different threshold value and a larger target score 

% Also, a higher target on level 3 could have changed gameplay experience.

% We also noticed that for some players, HR went down while playing the game. A correlation still needs to be understood that if stress can lower the HR as well.

% \paragraph{Future Work:} In the future, we plan to refine the game mechanics to explore whether a reward-based approach (`carrot') or a penalty-based approach (`stick') is more effective. One version of the game will decrease speed when players feel anxious and offer assistance, while another version will increase speed, adding difficulty as a form of penalty. Finding the player's intrinsic motivation to continue the gameplay while keeping their HR in control will be a crucial factor to discover in the future. Additionally,

\paragraph{Future Work} In the future, we aim to measure HRV and brainwave activity along with heart rate to gain deeper insights into stress levels of players during gameplay. We want to analyze the correlation between HR-based and HRV-based based stress levels. If stress breaches a threshold, the difficulty will be reduced. We also want to explore if introduction of other gamplay mechanics and elements helps players reduce stress while enhancing PX. We plan to put higher target score to see if PX stays high with a slightly longer gameplay time and also if HR adaptive levels helps players gain mastery faster. If significant differences found in mastery, HR-based adaptation can be used in the training phase of games or learning environments.

% We want to analyze the accuracy of HR-based stress levels by comparing with HRV and brainwave activity based stress levels.

\section{Conclusion}

% In this paper, we replicated the game Flappy Bird with the feature to control the gameplay speed using the players' real-time HR. The key goal was to investigate whether the players can remain calm within an increasingly difficult game environment. 

% Our additional research question was to analyze the overall gameplay experience under the circumstances where the game helps the players calm down by decreasing difficulty when they feel anxious. With a controlled experiment, our result demonstrated that players do enjoy the gameplay while the game action is controlled by their HR with an open-ended target.

% In this paper, we replicated the game Flappy Bird with the feature to control the gameplay speed using the player's real-time HR. The key goal was to analyze the overall gameplay experience under the circumstances where the game helps the players reduce stress by decreasing difficulty when they feel stressed. With a controlled experiment, our result demonstrated that players do enjoy the gameplay while the game action is controlled by their HR. Additionally, specific target inside the game can elicit more positive emotions. However, further investigation is needed to understand if HR-based adaptation can be used to reduce stress during gameplay while enhancing gameplay experience. 

In this paper, we replicated the game Flappy Bird with the feature to control the gameplay speed using the player's real-time HR. The key goal was to analyze the overall gameplay experience when game difficulty is reduced if the player feels stressed. With a controlled experiment, our result demonstrated that players do enjoy the gameplay while the game action is controlled by their HR. Additionally, specific targets inside the game can elicit more positive emotions. However, further investigation is needed to understand if HR-based adaptation can be used to reduce stress during gameplay while enhancing gameplay experience. 

% was to investigate how players react to HR controlled gameplay. Our additional research question 
% role of target score inside game needs further investigation.

% with an open-ended target.

\bibliographystyle{IEEEtran} 
\bibliography{bibliography}

\end{document}